\documentclass[12pt]{article}

\usepackage{amsmath,amssymb}

\oddsidemargin  - 5mm
\evensidemargin - 5mm

\newcommand{\nn}{\nonumber}

\newcommand{\sot}{SO(3, \mathbb{R})}

\begin{document}

\title{\bf
Unconstrained $SU(2)$ Yang-Mills Theory with Topological Term
in the Long-Wavelength Approximation}

\author{
A. M. Khvedelidze $^{a,b}$,  \,\,
D. M. Mladenov    $^c$,      \,\,
H.-P. Pavel       $^{c,d}$,      \, and  \,
G. R\"opke        $^d$ \\[4mm]
$^a${\it A. Razmadze Mathematical Institute, Tbilisi, 380093, Georgia}\\[2mm]
$^b$ {\it Laboratory of Information Technologies},\\
{\it Joint Institute for Nuclear Research, 141980 Dubna, Russia}\\[2mm]
$^c$ {\it Bogoliubov Laboratory of Theoretical Physics},\\
{\it Joint Institute for Nuclear Research, 141980 Dubna, Russia}\\[2mm]
$^d$ {\it Fachbereich Physik der Universit\"at Rostock, D-18051 Rostock, Germany}}

\date{April 29, 2003}
\maketitle

\begin{abstract}
The Hamiltonian reduction of $SU(2)$ Yang-Mills theory
for an arbitrary $\theta$ angle to an unconstrained nonlocal
theory of a self-interacting positive definite symmetric $3\times 3$
matrix field $S(x)$ is performed.
It is shown that, after exact projection to a reduced phase space,
the density of the Pontryagin index remains a pure divergence,
proving the $\theta$ independence of the unconstrained theory obtained.
An expansion of the nonlocal kinetic part of the Hamiltonian
in powers of the inverse coupling constant and truncation to lowest order,
however, lead to violation of the $\theta$ independence of the theory.
In order to maintain this property on the level of the local approximate theory,
a modified expansion in the inverse coupling constant is suggested,
which for a vanishing $\theta$ angle coincides with the original expansion.
The corresponding approximate Lagrangian up to second order in derivatives
is obtained, and the explicit form of the unconstrained analogue of the
Chern-Simons current linear in derivatives is given.
Finally, for the case of degenerate field configurations
$S(x)$ with $\mbox{rank}\|S\| = 1$, a nonlinear $\sigma$-type model is obtained,
with the Pontryagin topological term reducing to the Hopf invariant of the
mapping from the three-sphere $\mathbb{S}^3$ to the unit two-sphere
$\mathbb{S}^2$ in the Whitehead form.

\vspace{0.5cm}
\end{abstract}

\newpage

\bigskip

%%%%%%%%%%%%%%%%%%%%%%%%%%%%%%%%%%%%% 1 %%%%%%%%%%%%%%%%%%%%%%%%%%%%%%%%%%%%%%%%%%%

\section{Introduction}

%%%%%%%%%%%%%%%%%%%%%%%%%%%%%%%%%%%%%%%%%%%%%%%%%%%%%%%%%%%%%%%%%%%%%%%%%%%%%%%%%%%

For a complete understanding of the low-energy quantum phenomena
of Yang-Mills theory, it is necessary to have a nonperturbative,
gauge invariant description of the underlying classical theory
including the $\theta$-dependent Pontryagin term \cite{JackiwRebbi}-\cite{Jackiw}.
Several representations of Yang-Mills theory in terms of local gauge
invariant fields
have been proposed \cite{GoldJack}-\cite{Majumdar} in recent decades,
implementing the Gauss law as a generator of small gauge transformations.
However, in dealing with such local gauge invariant fields
special consideration is needed when the topological term is included,
since it is the four-divergence of a current
changing under large gauge transformations. In particular, the
consistency of constrained and unconstrained formulations of gauge
theories with topological term requires us to verify that, after projection
to the reduced phase space, the classical equations of motion for
the unconstrained variables remain $\theta$ independent.\footnote{
%__________________________________________________________________________%
The question of consistency of the elimination of redundant variables
in theories containing both constraints and pure divergencies,
the so-called ``divergence problem,'' was analyzed for the first time
in the context of the canonical reduction of general relativity
by Dirac \cite{Dirac} and by Arnowitt, Deser, and Misner \cite{ADM}.
}
%__________________________________________________________________________%
Furthermore, the question of which trace the large gauge transformations
with a nontrivial Pontryagin topological index leave on the local gauge
invariant fields has to be addressed.

Having this in mind, in the present paper we extend our approach
\cite{KP,AHG,KMPR}, to constructing the unconstrained form of
$SU(2)$ Yang-Mills theory to the case when the topological term is included
in the classical action.
We generalize the Hamiltonian reduction of classical $SU(2)$ Yang-Mills
field theory to arbitrary $\theta$ angle
by reformulating the original degenerate
Yang-Mills theory as a nonlocal theory of a self-interacting
positive definite symmetric $3\times 3$ matrix field.
The consistency of the Hamiltonian reduction in the presence of the
Pontryagin term is demonstrated by constructing
the canonical transformation, well defined on the reduced phase space,
that eliminates the $\theta$ dependence of the classical equations of motion
for the unconstrained variables.

With the aim of obtaining a practical form of the nonlocal unconstrained
Hamiltonian, we perform an expansion in powers of the inverse coupling constant,
equivalent to an expansion in the number of spatial derivatives.
We find that a straightforward application of the
derivative expansion violates the principle of $\theta$ independence of the
classical observables. To cure this problem,
we propose to exploit the property of chromoelectromagnetic duality of
pure Yang-Mills theory,
symmetry under the exchange of the chromoelectric and -magnetic fields.
The electric and magnetic fields are subject to
dual constraints, the Gauss law and Bianchi identity, and
only when both are satisfied, are the
classical equations of motion $\theta$ independent.
Thus any approximation in resolving the Gauss law constraints
should be consistent with the Bianchi identity.
We show how to use the Bianchi identity to
rearrange the derivative expansion in such a way
that the $\theta$ independence is restored to all orders on the classical level.

In order to have a representation of the gauge invariant degrees of freedom
suitable for a study of the low-energy phase of Yang-Mills theory,
we perform a principal-axes transformation of the symmetric tensor field
and obtain the unconstrained Hamiltonian
in terms of the principal-axes variables in the lowest order in $1/g$.
Carrying out an inverse Legendre transformation
to the corresponding unconstrained Lagrangian,
we find the explicit form of the unconstrained analog of the Chern-Simons
current, linear in the derivatives.

Finally, we consider the case of degenerate symmetric field
configurations $S$ with $\mbox{rank}\|S(x)\|= 1$.
We find a nonlinear classical theory of a three-dimensional unit-vector
${\bf n}$ field interacting with a scalar field.
Using typical boundary conditions for the unit-vector field
at spatial infinity, the Pontryagin topological charge density
reduces to the Abelian Chern-Simons invariant density \cite{Jackiw}.
We discuss its relation to the Hopf number of the
mapping from the three-sphere $\mathbb{S}^3$ to the unit two-sphere
$\mathbb{S}^2$ in the Whitehead representation \cite{Whitehead}.
The Abelian Chern-Simons invariant is known from different areas in
physics, in fluid mechanics as ``fluid helicity,''
in plasma physics and magnetohydrodynamics  as ``magnetic helicity''
\cite{Woltier}-\cite{Saffman}.
In the context of four-dimensional Yang-Mills theory a connection between
non-Abelian vacuum configurations and certain Abelian fields with nonvanishing
helicity has been established already in \cite{JackiwPi,NairJackiw}.

The paper is organized as follows.
In Sec. 2 the $\theta$ independence of classical Yang-Mills
theory in the framework of the constrained Hamiltonian formulation is revised.
Section 3 is devoted to the derivation of
unconstrained $SU(2)$ Yang-Mills theory for arbitrary $\theta$ angle.
The consistency of our reduction procedure is demonstrated by
explicitly quoting the canonical transformation,
which removes the $\theta$ dependence from the unconstrained
classical theory.
In Sec. 4 the unconstrained Hamiltonian up to
order $o(1/g)$ is obtained.
Section 5 presents the long-wavelength classical
Hamiltonian in terms of principal-axes variables.
The corresponding Lagrangian up to second order in derivatives,
and the unconstrained analogue of the Chern-Simons current, linear
in the derivatives, are obtained.
In Sec. 6 the unconstrained action for degenerate field configurations
is considered.
Section 7 finally gives our conclusions.
Several more technical details are presented in the Appendixes A, B, C, and D.
Appendix A summarizes our notation and definitions,
Appendix B is devoted to the question of the existence of the ``symmetric
gauge,'' in Appendix C the proof of the $\theta$-dependence of the
``naive'' $1/g$ approximation is given, and Appendix D contains some technical
details for the representation of the unconstrained theory in terms of
principal-axes variables.

%%%%%%%%%%%%%%%%%%%%%%%%%%%%%%%%%%%%%% 2 %%%%%%%%%%%%%%%%%%%%%%%%%%%%%%%%%%%%%%%

\section{Constrained Hamiltonian formulation}

%%%%%%%%%%%%%%%%%%%%%%%%%%%%%%%%%%%%%%%%%%%%%%%%%%%%%%%%%%%%%%%%%%%%%%%%%%%%%%%%

\label{sec:cedg}

Yang-Mills gauge fields are classified topologically by the
value of the Pontryagin index\footnote{
%________________________________________________________________________________%
The necessary notation and definitions for $SU(2)$ Yang-Mills
theory used in the text have been collected in Appendix A.
}
%________________________________________________________________________________%
\begin{equation}
p_1 = - \frac{1}{8 \, \pi^2} \, \int \, \mbox{tr} \, F \wedge F \,.
\end{equation}
Its density, the so-called topological charge density
$
Q\, = -\, (1/ 8 \, \pi^2)\, \mbox{tr} \, F \wedge F\,,
$
being locally exact
$ Q \, = \, d C, $
can be added to the conventional Yang-Mills Lagrangian
with arbitrary parameter $\theta$
\begin{equation}
\label{eq:Lagr}
{\cal L}  =  - \frac{1}{g^2}  \, \mbox{tr} \, F \wedge {}^\ast\! F -
\, \frac{\theta}{8\pi^2\, g^2} \, \mbox{tr} \, F \wedge F \,,
\end{equation}
without changing the classical equations of motion.
In the Hamiltonian formulation, this
shifts the canonical momenta,
conjugated to the field variables $A_{ai}$,
\begin{eqnarray} \label{eq:mom}
&&  \Pi_{ai} =
\frac{ \partial{\cal L} }{\partial {\dot A}_{ai} } =
{\dot A}_{ai} - \left( D_i (A) \right)_{ac} A_{c0} +
\frac{\theta}{8 \, \pi^2} \, B_{ai}\,,
\end{eqnarray}
by the magnetic field $(\theta /8 \, \pi^2) \, B_{ai}\,$.
As a result, the total  Hamiltonian \cite{DiracL,HenTeit}
of Yang-Mills theory with $\theta$ angle, as a functional
of canonical variables $(A_{a0}, \Pi_a)$ and  $(A_{ai}, \Pi_{ai})$
obeying the Poisson bracket relations
\begin{eqnarray} \label{eq:pbo}
&&
\{ A_{ai}(t, \vec{x})\, ,\Pi_{bj}(t, \vec{y}) \} =
\delta_{ab}\, \delta_{ij}\, \delta^{(3)}(\vec{x} - \vec{y}),\\
&&
\{ A_{a0}(t, \vec{x})\,,\Pi_{b}(t, \vec{y}) \} =
\delta_{ab}\, \delta^{(3)}(\vec{x} - \vec{y})
\end{eqnarray}
takes the form
\begin{equation} \label{eq:tothamn}
H_T = \int d^3 x \left[
\frac{1}{2}\left( \Pi_{ai} - \frac{\theta}{8\pi^2} B_{ai}\right)^2 +
\frac{1}{2}\, B_{ai}^2 - A_{a0} \left(D_i(A)\right)_{ac} \Pi_{ci}
+ \lambda_a \, \Pi_a
 \right]\,.
\end{equation}
Here, the linear combination of three primary constraints
\begin{equation}
\label{primconst}
\Pi_a (x) = 0
\end{equation}
with arbitrary functions $\lambda_a(x)$
and the secondary constraints, the non-Abelian Gauss law
\begin{equation}
 \label{eq:secconstr}
\left(D_i(A)\right)_{ac} \Pi_{ci} = 0 ,
\end{equation}
reflect the gauge invariance of the theory.

Based on the representation (\ref{eq:tothamn}) for the total Hamiltonian,
one can immediately verify that classical theories with different
value of the $\theta$ angle are equivalent.
Performing the canonical transformation
\begin{eqnarray}
A_{ai}(x)   &\longmapsto & A_{ai}(x), \nn\\
\Pi_{bj}(x) &\longmapsto & E_{bj}    :=
\Pi_{bj}(x) - \frac{\theta}{8 \pi^2}\, B_{bj}(x)
\label{eq:cantrtheta}
\end{eqnarray}
to the new variables $A_{ai}$ and $E_{bj}$,
and using the Bianchi identity
\begin{equation}
(D_i(A))_{ab} \, B_{bi}(A) = 0\,,
\end{equation}
one can then see that the $\theta$ dependence completely
disappears from the Hamiltonian (\ref{eq:tothamn}).
Note that the canonical transformation (\ref{eq:cantrtheta})
can be represented in the form
\begin{equation} \label{clctr}
E_{ai}\,  =
\, \Pi_{ai} - \, \theta \,\frac{\delta}{\delta A_{ai}}\, W[A]\,,
\end{equation}
where $W[A]$ denotes the winding number functional
\begin{equation}\label{clctr1}
W[A] \, = \, \int d^3 x \, K^0[A]
\end{equation}
constructed from the zero component of the Chern-Simons current
\begin{equation}
\label{CSC}
K^\mu[A] \, = - \frac{1}{16 \, \pi^2}\varepsilon^{\mu\alpha\beta\gamma}\,
\mbox{tr}
\left(
F_{\alpha\beta} \, A_\gamma - \frac{2}{3}\, A_\alpha \, A_\beta \, A_\gamma
\right).
\end{equation}
The question now arises whether, after reduction of Yang-Mills theory
including the topological term to the unconstrained system,
a transformation analogous to Eq. (\ref{eq:cantrtheta}) can be found
that correspondingly eliminates any $\theta$ dependence on the
reduced level, proving the consistency of the Hamiltonian reduction.

%%%%%%%%%%%%%%%%%%%%%%%%%%%%%%%%%%%%%% 3 %%%%%%%%%%%%%%%%%%%%%%%%%%%%%%%%%%%%%%%

\section{Unconstrained Hamiltonian formulation}

%%%%%%%%%%%%%%%%%%%%%%%%%%%%%%%%%%%%%%%%%%%%%%%%%%%%%%%%%%%%%%%%%%%%%%%%%%%%%%%%
\label{sec:unhf}

%#################################### 3.1 #################################################%

\subsection{Hamiltonian reduction for arbitrary $\theta$ angle}

%##########################################################################################%
\label{sec:unhf1}

In order to derive the unconstrained form of $SU(2)$ Yang Mills theory
with the $\theta$ angle we follow the method developed in \cite{KP}.
We perform the point transformation
\begin{equation}
\label{eq:gpottr}
A_{ai} \left(q, S \right) =
O_{ak}(q)\, S_{ki} +
\frac{1}{2 g} \,\varepsilon_{abc}\left(\partial_i O (q)\, O^T (q) \right)_{bc}
\end{equation}
from the gauge fields $A_{ai}(x)$ to the new set of three
fields $q_j(x),\,j = 1,2,3,$ parametrizing an orthogonal $3 \times 3$ matrix
$O(q)$ and the six fields $S_{ik}(x) = S_{ki}(x),\, i,k = 1,2,3,$
collected in the positive definite symmetric $3 \times 3$
matrix $S(x)$.\footnote{
%____________________________________________________________________________%
It is necessary to note that a decomposition similar to Eq. (\ref{eq:gpottr})
was used in \cite{Simonov} as a generalization of the well-known polar
decomposition valid for arbitrary quadratic matrices.}
%_____________________________________________________________________________%
Equation (\ref{eq:gpottr}) can be seen as a gauge transformation to the
new field configuration $S(x)$ which satisfies the ``symmetric gauge''
condition
\begin{equation}
\label{symgauge}
\chi_a(S):= \varepsilon_{abc}\, S_{bc} = 0\,.
\end{equation}
The complete analysis of the existence and uniqueness of this gauge,
i.e., whether any gauge potential $A_{ai}$ can be made symmetric by a
unique gauge transformation, is a complex mathematical problem.
Here we shall consider the transformation (\ref{eq:gpottr}) in a region where
the uniqueness and regularity of the change of coordinates
can be guaranteed.
In Appendix B, we prove the existence and uniqueness of the symmetric
gauge for the case of a nondegenerate matrix $A$
using the inverse coupling constant expansion.
Furthermore, as an illustration of the obstruction of the uniqueness of
the symmetric gauge fixing (the appearance of Gribov copies) for degenerate
matrices $A$, the Wu-Yang monopole configuration is considered.
Although it is antisymmetric in space and color indices,
it can be brought into the symmetric form, but there exist two
gauge transformations by which this can be achieved.
The case of a degenerate matrix field $S$, $\det\|S\|=0$, will be discussed
for the special situation $\mbox{rank}\|S\|=1$ in Sec. 6.

The transformation (\ref{eq:gpottr}) induces a point canonical
transformation linear in the new momenta $P_{ik}(x)$ and $p_i(x)$,
conjugated with $S_{ik}(x)$ and $q_i (x)$, respectively.
Their expressions in terms of the old variables
$(A_{ai}(x)\,, \Pi_{ai}(x))$ can be obtained from the requirement
of the canonical invariance of the symplectic one-form
\begin{equation}
\label{eq:can1f}
\sum^3_{i, a = 1 }\, \Pi_{ai}\, \dot{A}_{ai}\, dt  =
\sum^3_{i, j = 1}\, P_{ij}\, \dot{S}_{ij}\, dt  +
\,\sum^3_{i = 1} \, p_i\, \dot{q}_i\, dt
\end{equation}
with the fundamental brackets
\begin{eqnarray}
\label{eq:Diracb}
&&
\{ S_{i j}(t, \vec{x}) \,, P_{k l}(t, \vec{y}) \} = \frac{1}{2}\,
\left(\delta_{i k}\, \delta_{j l} + \delta_{i l}\, \delta_{j k} \right)\,
\delta^{(3)}(\vec{x} - \vec{y}),\\
&&
\{ q_i(t, \vec{x}) \,,  p_j (t, \vec{y})\} =
\delta_{i j}\, \delta^{(3)}(\vec{x} - \vec{y})
\end{eqnarray}
for the new canonical
pairs $\left(S_{ij}(x)\,, P_{ij}(x)\right)$ and
$\left(q_{i}(x)\,, p_{i}(x)\right)$.
The brackets (\ref{eq:Diracb}) account for the second-class symmetry constraints
$S_{ij} = S_{ji}$ and $P_{ij} = P_{ji}$ and therefore are Dirac brackets.
As a result we obtain the expression
\begin{equation}
\label{eq:elpotn}
\Pi_{ai} = O_{ak}(q) \biggl[\,
P_{\ ki} + g \, \varepsilon _{kin}{}^\ast\! D^{-1}_{nm}(S)
\left( {\cal S}_m - \Omega^{-1}_{jm} p_j\right) \,
\biggr]
\end{equation}
for the old momenta $\Pi_{ai}$ in terms of the new canonical variables
(for a detailed derivation see \cite{KP}).
Here ${}^\ast\! D_{mn}^{-1}(S)$ denotes the inverse of
the differential matrix operator\footnote{
%-----------------------------------------------------------------------
Note that the operator ${}^\ast\! D_{mn}(S)$
 corresponds in the conventional gauge-fixing
method to the so-called Faddeev-Popov (FP) operator,
the matrix of Poisson brackets between the Gauss law constraint
(\ref{eq:secconstr}) and  the symmetric gauge (\ref{symgauge}),
 $\{ \left(D_i(S)\right)_{mc} \Pi_{ci}(x), \chi_n (y)\}=
 {}^\ast\! D_{mn}(S)\delta^3(x-y).$ }
%----------------------------------------------------------------
\begin{equation}
\label{DeltaQ}
{}^\ast\! D_{mn}(S)  =  \varepsilon_{njc}\left(D_j(S)\right)_{mc}\,,
\end{equation}
the vector ${\cal S}$ is defined as
\begin{equation}
{\cal S}_m   = \frac{1}{g} \, (D_j(S))_{mn}P_{nj}~,
\end{equation}
and the matrix $\Omega^{-1}$ is the inverse of
\begin{equation}
\label{Omega}
\Omega_{ni}(q) \, : =
\,-\frac{1}{2}\, \varepsilon_{nbc}\,
\left( O^T(q)\, \frac{\partial O(q)}{\partial q_i}\right)_{bc}.
\end{equation}

Here we would like to comment on the geometrical meaning of the above expressions.
The vector ${\cal S} $ coincides up to divergence with the spin density part
of the Noetherian angular momentum after projection to the surface given by
the Gauss law constraints.
Furthermore, the matrix $\Omega^{-1}$ defines the main geometrical
structures on the $SO(3,R)$ group manifold, namely, the three
left-invariant Killing vector fields
$\eta_a :=\Omega^{-1}_{ja} \partial/\partial q_j$ obeying the
$so(3)$ algebra $[ \eta_a, \eta_b]= \epsilon_{abc}\eta_c$,
and the invariant  metric $g:=-\mbox{tr}\left(O^TdO O^TdO\right)=
(1/2)\left(\Omega^T\Omega\right)_{ij}dq_idq_j$
as the standard metric on $S^3$.
Since $\det\Omega$ is proportional to the Haar measure
on $SO(3, R)$ $\sqrt{\det \|g\|} =|\det\|\Omega(q)\||$,
and it is expected to vanish at certain coordinate singularities
(see also, e.g., the  discussion in  chap. 8 of \cite{Creutz}).
In deriving the expression (\ref{eq:elpotn}) we shall here limit ourselves to
the region where the matrix $\Omega$ is invertible.

The main advantage of introducing the variables $S_{ij}$ and $q_i$
is that they Abelianize the non-Abelian Gauss law constraints
(\ref{eq:secconstr}).
In terms of the new variables the Gauss law constraints
\begin{equation}
\label{Omega-1}
g \, O_{as}(q)\, \Omega^{-1}_{\ is}(q)\, p_i = 0
\end{equation}
depend only on $(q_i,p_i)$, showing
that the variables $(S_{ij} \,, P_{ij})$ are gauge-invariant, physical fields.
Hence, assuming $\det\Omega(q)\ne 0$\
in Eqs. (\ref{eq:elpotn}) and (\ref{Omega-1}),
the reduced Hamiltonian, defined as the projection of the
total Hamiltonian onto the constraint shell,
can be obtained from Eq.(\ref{eq:tothamn}) by
imposing the equivalent set of Abelian constraints
\begin{equation}
\label{p_a=0}
 p_i = 0\,.
\end{equation}
Due to gauge invariance, the reduced Hamiltonian is
independent of the coordinates $q_i$  canonically conjugated
to $p_i$ and is hence a function  of the unconstrained
gauge-invariant variables $S_{ij}$ and $P_{ij}$ only
\begin{equation}
\label{eq:uncYME}
H = \int d^3{x}\, \biggl[\,
\frac{1}{2}\, \left( P_{ai} - \frac{\theta}{8 \pi^2}\, B^{(+)}_{ai}(S)
\right)^2 +
\left( P_a - \frac{\theta}{8 \pi^2}\, B^{(-)}_{a}(S)\right)^2 +
\frac{1}{2}\, V(S)\,
\biggr]\,.
\end{equation}
Here the $P_a$ denotes the nonlocal functional,
according to Eq. (\ref{eq:elpotn}) defined as the solution of the system of
differential equations
\begin{equation}
\label{vecE}
{}^\ast\! D_{ks}(S) P_s = (D_j(S))_{kn}P_{nj}\,.
\end{equation}
The nonlocal second term in the Hamiltonian (\ref{eq:uncYME})
therefore stems from the antisymmetric part of the
$\Pi_{ai}$, which remains after implementing Gauss's law $p_a = 0$,
in terms of the physical $P_{ai}$. Hence this term contains $FP^{-2}$,
[see Eq. (26)], and is the analogue of the
well-known nonlocal part of the Hamiltonian in the Coulomb gauge
(see, e.g., \cite{ChrLee}).

Furthermore,
\begin{equation}
\label{symasymB}
B^{(+)}_{ai}(S) := \frac{1}{2}\,[B_{ai}(S) + B_{ia}(S)]\,, \qquad
B^{(-)}_a (S):= \frac{1}{2}\,\varepsilon_{abc} \, B_{bc}(S)
\end{equation}
denote the symmetric and antisymmetric
parts of the reduced chromomagnetic field
\begin{equation}
\label{redB}
B_{ai}(S) = \varepsilon_{ijk}\,
\left(\partial_j S_{ak} + \frac{g}{2}\,
\varepsilon_{abc}\, S_{bj}\, S_{ck}\right)\,.
\end{equation}
It is the same functional of the symmetric field $S$ as the original
$B_{ai}(A)$, since the chromomagnetic field transforms homogeneously
under the change of coordinates (\ref{eq:gpottr}).
Finally, the potential $V(S)$ is the square of the reduced magnetic field
(\ref{redB}),
\begin{equation}
\label{V(S)}
V(S)\ d^3x = B_{ai}^2(S)\ d^3x =
\frac{1}{2} \ \mbox{tr}\,{}^\ast\! F^{(3)}\,\wedge \, F^{(3)},
\end{equation}
with the curvature two-form in three-dimensional Euclidean space
\begin{equation}
\label{3F}
 F^{(3)} = d S + S\, \wedge\, S\,,
\end{equation}
in terms of the symmetric one-form
\begin{equation}
\label{3S}
S = g\tau_k\, S_{kl} \, dx_l, \qquad k, l = 1,2,3\,,
\end{equation}
whose six components depend on the time variable as an external parameter.
The reduced chromomagnetic field (\ref{redB}) is given in terms of the
dual field strength ${}^\ast\! F^{(3)}$ as
$B_{ai}(S) =\frac{1}{2}\, \varepsilon_{ijk}\, F^{(3)}_{\ ajk}\,$.

%#################################### 3.2 ##################################%

\subsection{Canonical equivalence of unconstrained theories with \\
 different $\theta$ angles}

%############################################################################%
\label{subsecIIB}

For the original degenerate action in terms of the $A_{\mu}$ fields
the equivalence of classical theories
with arbitrary values of $\theta$ angle has been reviewed in
Sec. \ref{sec:cedg}.
Let us now examine the same problem for the unconstrained
theory derived considering the analogue of the canonical transformation
(\ref{eq:cantrtheta}) after projection onto the constraint surface,
\begin{eqnarray}
S_{ai}(x) & \longmapsto & S_{ai}(x), \nn\\
P_{bj}(x) & \longmapsto & {\cal E}_{bj}(x) :=
P_{bj}(x) - \frac{\theta}{8\,\pi^2}\, B^{(+)}_{bj}(x)\label{eq:uncantrtheta}.
\end{eqnarray}
One can easily check that this transformation
to new variables $S_{ai}$ and ${\cal E}_{bj}$
is canonical with respect to the Dirac brackets (\ref{eq:Diracb}).
In terms of the new variables $S_{ai}$ and ${\cal E}_{bj}$ the
Hamiltonian (\ref{eq:uncYME}) can be written as
\begin{equation}
\label{eq:ht1}
H =
\int d^3 x
\biggl[\,
\frac{1}{2}\, {\cal E}_{ai}^2 + {\cal E}_{a}^2 + \frac{1}{2}\, V(S) \,
\biggr]\,,
\end{equation}
with ${\cal E}_a$  defined as
\begin{equation}
{\cal E}_a := P_a - \frac{\theta}{8\, \pi^2} \, B^{(-)}_{a}\,.
\end{equation}
Now, if $P_a$ is a solution of Eq. (\ref{vecE}), then
${\cal E}_a$ is a solution of the same equation
\begin{equation}
\label{vecE"}
{}^\ast\! D_{ks}(S){\cal E} _s = (D_j(S))_{kn}{\cal E}_{nj}
\end{equation}
with the replacement $P_{ai} \longmapsto {\cal E}_{ai}$,
since the reduced field $B_{ai}$ satisfies the Bianchi identity
\begin{equation}
\label{BI}
(D_i(S))_{ab} \, B_{bi}(S) = 0\,.
\end{equation}
Hence we arrive at the same unconstrained Hamiltonian system
(\ref{eq:ht1}) and (\ref{vecE"}) with vanishing $\theta$ angle.
Note that after the elimination of the three unphysical fields $q_j(x)$
the projected canonical transformation (\ref{eq:uncantrtheta})
that removes the $\theta$ dependence from the Hamiltonian can be written as
\begin{equation}
{\cal E}_{bj}(x) =
P_{bj}(x) - \theta \, \frac{\delta}{\delta S_{bj}}\, W[S]\,,
\end{equation}
which is of the same form as Eq. (\ref{clctr}) with the nine gauge fields
$A_{ik}(x)$ replaced by the six unconstrained fields $S_{ik}(x)$.

In summary, the exact projection to a reduced phase space leads to an
unconstrained system
whose equations of motion are consistent with the original degenerate theory
in the sense that they are $\theta$ independent.
Thus if our consideration is restricted only to
the classical level of the exact nonlocal unconstrained theory, the
generalization to arbitrary $\theta$ angle can be avoided.\footnote{
%______________________________________________________________________________%
The extension of the proof of $\theta$ independence to
quantum theory requires showing the unitarity of
the operator corresponding to the transformation
(\ref{eq:uncantrtheta}).}
%______________________________________________________________________________%
However, in order to work with such a complicated nonlocal
Hamiltonian it is necessary to make approximations, such as, for example,
expansion in the number of spatial derivatives, which we shall carry
out in the next section.
For these one has to check that this approximation
is free of the ``divergence problem", that is, all terms in the corresponding
truncated action containing the $\theta$ angle can be collected into a
four-divergence and all dependence on $\theta$
disappears from the classical equations of motion.

%%%%%%%%%%%%%%%%%%%%%%%%%%%%%%%%%%%%%% 4 %%%%%%%%%%%%%%%%%%%%%%%%%%%%%%%%%%%%%%%

\section{Expansion of the unconstrained Hamiltonian in $1/g$}

%%%%%%%%%%%%%%%%%%%%%%%%%%%%%%%%%%%%%%%%%%%%%%%%%%%%%%%%%%%%%%%%%%%%%%%%%%%%%%%%

\label{secIV}

Let us now consider the regime when the unconstrained
fields are slowly varying in space-time and expand
the nonlocal part of the kinetic term in the unconstrained
Hamiltonian (\ref{eq:uncYME}) as a series of terms with increasing
powers of the inverse coupling constant $1/g$,
equivalent to an expansion in the
number of spatial derivatives of field and momentum.
Our expansion is purely formal and we shall not
study the question of its convergence in this work.
We shall see that for nonvanishing $\theta$ angle
a straightforward expansion in $1/g$ leads to the above mentioned
``divergence problem," and suggest an
improved form of the expansion in $1/g$ of the unconstrained Hamiltonian
exploiting the Bianchi identity.

%#################################### 4.1 ####################################%

\subsection{Divergence problem in lowest-order approximation}

%##############################################################################%
\label{Sec:IV1}

According to \cite{KP}, the nonlocal funtional $P_a$ in the
unconstrained Hamiltonian (\ref{eq:ht1}), defined as a solution of the system of
linear differential equations (\ref{vecE}), can formally be expanded
in powers of $1/g$.
The vector $P_a$ is then given as a sum of terms containing an increasing
number of spatial derivatives of field and momentum
\begin{equation}
\label{vecEmexp}
P_s (S, P) = \sum_{n=0}^{\infty}(1/g)^n\, a_s^{(n)}(S, P).
\end{equation}
The zeroth-order term is
\begin{equation}
\label{vecE1}
a^{(0)}_{s} =
\gamma^{-1}_{sk}\varepsilon_{klm}\left(PS\right)_{lm}\,,
\end{equation}
with $\gamma_{ik}:= S_{ik} - \delta_{ik}\, \mbox{tr}\, S$,
and the first-order term is determined as
\begin{equation}
\label{vecE2}
a^{(1)}_{s} = - \, \gamma^{-1}_{sl}\,
\left[
(\mbox{rot}\ \vec{a}^{(0)})_l + \partial_k P_{kl}
\right]
\end{equation}
from the zeroth-order term.
The higher terms are then obtained by the simple recurrence relations
\begin{equation} \label{vecE3}
a^{(n+1)}_{s} =
- \,\gamma^{-1}_{sl}(\mbox{rot}\, {\vec {a}}^{\ (n)})_l \,.
\end{equation}
Inserting these expressions into Eq. (\ref{eq:uncYME})
we obtain the corresponding expansion of the unconstrained Hamiltonian as
a series in higher and higher numbers of derivatives.

Let us check whether the truncation of the expansion
(\ref{vecEmexp}) to lowest order is consistent with $\theta$ independence,
that is, whether all $\theta$-dependent
terms can be collected into a total four-divergence after Legendre
transformation to the corresponding Lagrangian.
In $o(1/g)$ approximation (\ref{vecE1}), the Hamiltonian reads\footnote{
%--------------------------------
When all spatial derivatives of the
fields and momenta are neglected, Yang-Mills theory reduces to the so-called
Yang-Mills mechanics and its $\theta$ independence has been shown
in \cite{AHG}.
%-------------------------------
}
\begin{equation}
\label{eq:ham2}
H^{(2)} =
\int d^3{x}\,
\biggl[\,
\, \frac{1}{2}\mbox{tr}\,\left(P - \frac{\theta}{8 \pi^2}\, B^{(+)}\right)^2
+\left(a_s^{(0)}(S, P) - \frac{\theta}{8\, \pi^2}\, B^{(-)}_s
\right)^2 + \,\frac{1}{2} V(S) \ \biggr]\,,
\end{equation}
where $B^{(+)}$ and $B^{(-)}$ denote the symmetric and antisymmetric
parts of the chromomagnetic field, defined in Eq. (\ref{symasymB}).

After inverse Legendre transformation of the Hamiltonian (\ref{eq:ham2}),
the $\theta$-dependent terms in the corresponding Lagrangian
cannot be collected into a total four-divergence, as is shown in Appendix C,
and therefore contribute to the unconstrained equations of motion.
Hence, on applying a straightforward derivative expansion to the
Yang-Mills theory with a topological term after projection to a reduced phase
space, we face the ``divergence problem" dicussed above.

%#################################### 4.2 ####################################%

\subsection{Improved $1/g$ expansion using the Bianchi identity}

%##############################################################################%
\label{SECTIONIV2}

In order to avoid the ``divergence problem''  one can proceed as follows.
Let us consider additionally to the differential equation (\ref{vecE}),
which determines the nonlocal term $P_a$, the Bianchi identity (\ref{BI})
as an equation for determination of the antisymmetric part $B^{(-)}_s$
of the chromomagnetic field
\begin{equation}
\label{eq:abm}
^{\ast}\! D_{ks}(S)\, B^{(-)}_s = (D_i(S))_{kl}\, B^{(+)}_{li}
\end{equation}
in terms of its symmetric part $B^{(+)}_{bc}$.
The complete analogy of this equation with Eq. (\ref{vecE}) expresses the
duality of the chromoelectric and chromagnetic fields on the unconstrained level.
Hence one can write
\begin{equation}
\label{eq:BG}
^{\ast}\! D_{ks}(S)\,
\left[P_s - \frac{\theta}{8 \pi^2}\, B^{(-)}_s \right] =
(D_i(S))_{kl}\,
\left[ P_{li} - \frac{\theta}{8\, \pi^2}\, B^{(+)}_{li}\right].
\end{equation}
Using the same type of spatial derivative expansion as before in
Eqs. (\ref{vecE1})-(\ref{vecE3}), we obtain
\begin{equation}
\label{P-B-}
P_s - \frac{\theta}{8\, \pi^2}\, B^{(-)}_s =
\sum_{n = 0}^{\infty}\,(1/g)^n\,
a^{(n)}_s\left(S, P - \frac{\theta}{8 \pi^2} \, B^{(+)}\right)\,.
\end{equation}
In this way we achieve a form of the derivative expansion such that the
unconstrained Hamiltonian is a functional of the
field combination $P_{ai} - (\theta/8 \,\pi^2)\, B^{(+)}_{ai}$,
\begin{equation}
\label{eq:uncYMEi}
H =
\int d^3{x}\, \biggl\{\,
\frac{1}{2}\, \left(P_{ai} - \frac{\theta}{8\pi^2}\, B^{(+)}_{ai}\right)^2 +
\left[
\sum_{n=0}^{\infty}(1/g)^n\,
a^{(n)}_i\left(S, P - \frac{\theta}{8 \pi^2}\, B^{(+)}\right)\right]^2
+ \frac{1}{2}\, V(S) \, \biggr\}\,,
\end{equation}
explicitly showing the chromoelectromagnetic duality on the reduced level
and hence free of the ``divergence problem".
To obtain the unconstrained Hamiltonian up to leading order $o(1/g)$,
only the lowest term $a^{(0)}_s(S, P - (\theta/8 \,\pi^2)\, B^{(+)})$
in the sum in Eq. (\ref{eq:uncYMEi}) has to be taken into account, so that
\begin{equation}
\label{eq:iham2}
H^{(2)} =
\frac{1}{2}\int d^3{x}\, \biggl\{\,
\, \mbox{tr}\, \left( P - \frac{\theta}{8\pi^2}\, B^{(+)}\right)^2 -
\frac{1}{ \det^2 \gamma}\, \mbox{tr}\,
\left(\gamma\, \left[S, P - \frac{\theta}{8 \,\pi^2}\, B^{(+)}\right]\,
 \gamma\right)^2 + \, V(S)\biggr\}\,.
\end{equation}
The advantage of this Hamiltonian compared with Eq. (\ref{eq:ham2}), derived
before, is that the classical equations of motion following from
Eq. (\ref{eq:iham2}) are $\theta$ independent.
In order to obtain a transparent
form of the corresponding surface term in the unconstrained action,
it is useful to perform a principal-axes transformation of the symmetric
matrix field $S(x)$.

%%%%%%%%%%%%%%%%%%%%%%%%%%%%%%%%%%%%%% 5 %%%%%%%%%%%%%%%%%%%%%%%%%%%%%%%%%%%%%%%

\section{Long-wavelength approximation to reduced theory}

%%%%%%%%%%%%%%%%%%%%%%%%%%%%%%%%%%%%%%%%%%%%%%%%%%%%%%%%%%%%%%%%%%%%%%%%%%%%%%%%

\label{sec:V}

In this section we shall first rewrite the
unconstrained Hamiltonian (\ref{eq:iham2}) in terms of principal-axes variables
of the symmetric tensor field $S_{ij}$.
The corresponding second-order Lagrangian $L^{(2)}$ is then obtained
via Legendre transformation and the form of the corresponding
unconstrained total divergence derived in an explicit way.

%#################################### 5.1 ####################################%

\subsection{Hamiltonian in terms of principal-axes variables}

%##############################################################################%

\label{secIV2}

In \cite{KP} it was shown that the field $S_{ij}(x)$
transforms as a second-rank tensor under spatial rotations.
This can be used to explicitly separate the rotational degrees of
freedom from the scalars in the Hamiltonian (\ref{eq:iham2}).
Following \cite{KP}, we introduce  the principal-axes representation
of the  symmetric $3 \times 3$ matrix field $S(x)$,
\begin{equation}
\label{eq:mainax}
S (x) =
R^T[\chi(x)]
\left(
\begin{array}{ccc}
\phi_1(x)  &   0         &    0       \\
0          & \phi_2(x)   &    0        \\
0          &   0         &   \phi_3(x)
\end{array}
\right )
R[\chi(x)]\,.
\end{equation}
The Jacobian of this transformation is
\begin{equation}
J\left(\frac{S_{ij}[\phi, \chi]}{\phi_{k}, \chi_{l}}\right) \propto
\prod_{i\neq j}\mid \phi_i(x) - \phi_j(x) \mid,
\end{equation}
and thus Eq. (\ref{eq:mainax}) can be used as a definition of
the new configuration variables,
the three diagonal fields $\phi_1, \phi_2, \phi_3$ and
the three angular fields $\chi_1, \chi_2,\chi_3$,
only if all eigenvalues of the matrix $S$ are different.
To have uniqueness of the inverse transformation we assume here
that
\begin{equation}
\label{princorb}
0 < \phi_1(x) < \phi_2(x) < \phi_3(x)\,.
\end{equation}
The variables $\phi_i$ in the principal-axes transformation (\ref{eq:mainax})
parametrize the orbits of the action of a group element $g\in\sot$ on symmetric
matrices $S\rightarrow S^\prime = g\, S\, g^{-1} $.
The configuration (\ref{princorb}) belongs to the so-called  principal orbit
class,
whereas all orbits with coinciding eigenvalues of the matrix $S$ are
singular orbits \cite{ORaf}.
In order to parametrize configurations belonging to a singular stratum
one should in principle use a decomposition of the $S$ field different from the
above principal-axes transformation (\ref{eq:mainax}).
Alternatively, one can consider the singular orbits as the boundary of the
principal orbit-type stratum and study the corresponding dynamics using a
certain limiting procedure.
\footnote{
%--------------------------------------------------------------------
The relation between an explicit parametrization
of the singular strata and their description as a certain limit
of the principal orbit stratum has been studied recently in \cite{AD}
investigating the geodesic motion on the $GL(n, R)$ group manifold.}
%------------------------------------------------------------------------
In this section we shall limit ourselves to the consideration of the dynamics
on the principal orbits and leave the important case of the singular orbits
expected to contain interesting physics for future studies.

The momenta $\pi_i$ and $p_{\chi_i}$, canonically conjugate
to the diagonal elements $\phi_i$ and  $\chi_i$, can be found using
the condition of the canonical invariance of the symplectic one-form
\begin{equation}
\sum^3_{i,j = 1}\, P_{ij}\, \dot{S}_{ij}\, dt  =
\sum^3_{i = 1}\, \pi_{i}\, \dot{\phi}_{i} dt  +
\sum^3_{i = 1}\, p_{\chi_i}\, \dot{\chi}_i  dt\,.
\end{equation}
The original physical momenta $P_{ik}$, expressed
in terms of the new canonical variables, read
\begin{eqnarray} \label{eq:newmom}
P(x) =
R^T(x)\,
\sum_{s = 1}^3\left(
\pi_s(x) \, \overline{\alpha}_s + \frac{1}{2}{\cal P}_s(x) \,\alpha_s
\right)\,
R(x)\,.
\end{eqnarray}
Here $\overline{\alpha}_i$ and $\alpha_i$
denote the diagonal and off-diagonal basis elements for symmetric matrices
with the orthogonality relations
%\label{eq:symb}
$\mbox{tr}\, (\overline{\alpha}_i \, \overline{\alpha}_j) = \delta_{ij},$
$\mbox{tr}\, ({\alpha}_i \, {\alpha}_j) = 2\, \delta_{ij},$
$\mbox{tr}\, (\overline{\alpha}_i \, {\alpha}_j) = 0,$
and
\begin{equation}
\label{P+approx}
{\cal P}_i (x) = - \, \frac{\xi_i(x)}{\phi_j(x) - \phi_k(x)}
\qquad
(\mbox{cyclic permutations} \,\, i\not=j\not= k )\,.
\end{equation}
The $\xi_i$ are the three $\sot$ right-invariant Killing vector fields,
satisfying locally the ``intrinsic frame" angular momentum brackets
$\{\xi_i(x),\xi_j(y)\}=-\epsilon_{ijk}\xi_k(x)\delta(x-y)$, and are
given in terms of the angles $\chi_i$ and their conjugated momenta
$p_{\chi_i}$ via\footnote{
%---------------------------
In terms of the Euler angles $\chi_i=(\alpha,\beta,\gamma)$ the three
right-invariant Killing vector fields $\xi_i$ read
$\xi_1 = \sin\gamma\, p_\alpha - (\cos\gamma/\sin\alpha)\, p_\beta +
          \cos\gamma\cot\alpha\, p_\gamma$,
$\xi_2 = \cos\gamma\, p_\alpha + (\sin\gamma/\sin\alpha)\, p_\beta -
          \sin\gamma\cot\alpha\, p_\gamma$, and
$\xi_3 = p_\gamma$.
%---------------------------
}
\begin{equation}
\xi_i =\sum_{j=1}^{3} M^{-1}_{ji} p_{\chi_j}\,,
\end{equation}
where the matrix $M$ is
\begin{equation} \label{eq:MCr}
M_{ji} := - \frac{1}{2}\, \sum_{a,b=1}^{3}\varepsilon_{jab}
\left(\frac{\partial R}{\partial \chi_i}\, R^T\right)_{ab}\,.
\end{equation}

In terms of the principal-axes variables (\ref{eq:mainax}),
the $o(1/g)$ Hamiltonian (\ref{eq:iham2})
can be written in the form (for technical details see Appendix D)
\begin{equation}
\label{eq:unch2}
H^{(2)}  =
\frac{1}{2}\, \int d^3x
\left[
\sum_{i=1}^3
\left(\pi_i - \frac{\theta}{8\pi^2}\, {\beta}_i\right)^2  +
\sum_{cyclic}^{i,j,k}\, k_i \left(
\xi_i + \frac{\theta}{8\pi^2}\,(\phi_j - \phi_k)\,b_{i}\,\right)^2
+  V(\phi,\chi)
\right]\,,
\end{equation}
with
the diagonal components $\beta_i$ and
the off-diagonal components $b_i$ of the
the symmetric part of the chromomagnetic field (see Appendix D)
\begin{eqnarray}
\beta_i &=&
g\phi_j\phi_k - (\phi_i-\phi_j)\Gamma_{ikj} +
(\phi_i-\phi_k)\Gamma_{ijk}
\nonumber\\
&&(\mbox{cyclic permutations} \,\, i\not=j\not= k )\,,
\label{beta_i}
\\
b_i &=&
X_i(\phi_j-\phi_k)-(\phi_i-\phi_j)\Gamma_{ijj} +
(\phi_i-\phi_k)\Gamma_{ikk}
\nonumber\\
&&(\mbox{cyclic permutations} \,\, i\not=j\not= k ),
\label{b_i}
\end{eqnarray}
the abbreviations
\begin{equation}
\label{eq:km}
k_i := \frac{\phi_j^2 + \phi_k^2}{(\phi_j^2 - \phi_k^2)^2} \quad
(\mbox{cyclic permutations}\,\,  i\not = j\not = k )\,,
\end{equation}
and the potential $V$, defined in Eq. (\ref{V(S)}) and rewritten in the
principal-axes variables as (see \cite{KP} and the errata \cite{ErrKP})
\begin{equation}
\label{Vinhom1}
V(\phi,\chi) = \sum_{i\neq j}^{3}
\left[
 (\phi_i-\phi_j)\Gamma_{iij} - X_j \phi_i\right]^2
+ \sum_{cyclic}^{i,j,k}\left[(\phi_i-\phi_k)\Gamma_{ijk} -
(\phi_i - \phi_k)\Gamma_{ikj} - g\phi_j \phi_k\right]^2.
\end{equation}
The dependence on the angular variables $\chi_i$ in Eqs.(\ref{beta_i}),
(\ref{b_i}), and (\ref{Vinhom1}) has been collected into
the vector fields
\begin{equation}
X_i :=\sum_{j=1}^{3} R_{ij}\,\partial_j \,,
\end{equation}
and the components of the connection one-form $\Gamma$
\begin{equation}
\label{Gammaspace}
\Gamma_{aib} := \left(X_i R\, R^T \right)_{ab}.
\end{equation}
We see that through the principal-axes transformation of the symmetric tensor
field $S$, the highest order $g^2$ terms in the Hamiltonian (\ref{eq:unch2}),
which are proportional to the
spatially homogeneous part $V_{\rm hom}$ of the potential (\ref{Vinhom1}),
\begin{equation}
\label{Vhom}
V_{\rm hom} = g^2\left(\phi_1^2\phi_2^2+\phi_2^2\phi_3^2+
\phi_3^2\phi_1^2\right),
\end{equation}
depend only on the diagonal fields $\phi_i$, while
the rotational degrees of freedom $\chi_i$ and their canonically
conjugate momenta $p_{\chi_i}$ appear in the unconstrained Hamiltonian
(\ref{eq:unch2}) only via the Killing vector fields $\xi_i$,
the connection $\Gamma$, and the vectors $X_i$.

The transformation (\ref{eq:uncantrtheta}), rewritten
in terms of angular and scalar variables,
\begin{eqnarray}
\label{eq:crt}
&&
\pi_i \longmapsto \pi_i + \frac{\theta}{8\pi^2}\,\beta_i \,, \qquad
\phi_i \longmapsto \phi_i \,, \nn\\
&&
\xi_i \longmapsto \xi_i - \frac{\theta}{8\pi^2}\,(\phi_j - \phi_k)\,b_i \,,
\end{eqnarray}
excludes the $\theta$ dependence from the Hamiltonian (\ref{eq:unch2}),
reducing it to the zero $\theta$ angle expression \cite{KP}
\begin{equation}
\label{eq:uncz}
H^{(2)} \, = \,
\frac{1}{2}\, \int\, d^3x
\left[
\sum_{i=1}^3 \pi_i^2  +
\sum_{cyclic}^{i,j,k}\xi_i^2\frac{\phi_j^2+\phi_k^2}{(\phi_j^2-\phi_k^2)^2}
+V(\phi, \chi)
\right]\,.
\end{equation}

%#################################### 5.2 ######################################%

\subsection{Second-order unconstrained Lagrangian}

%###############################################################################%

We are now ready to derive the Lagrangian up to second order in derivatives
corresponding to the Hamiltonian (\ref{eq:unch2}).
Carrying out the inverse Legendre transformation,
\begin{eqnarray}
\dot{\phi}_i &=& \pi_i - \frac{\theta}{8\pi^2}\beta_i\,,\\
\label{eq:ilt}
\dot \chi_i &=&\sum_{j=1}^{3} G_{ij}\Big(p_{\chi_j} -
\frac{\theta}{8\pi^2}\sum_{cyclic}^{a,b,c}M^T_{ja}
(\phi_b -\phi_c)\,b_{a}\Big)\,,
\end{eqnarray}
with the matrix $M$ given in Eq. (\ref{eq:MCr})
and the $3\times 3$ matrix $G$,
\begin{equation}
G= M^{-1} k {M^{-1}}^T\,,
\end{equation}
similar to the diagonal matrix
$k=\mbox{diag}\|{k_1, k_2, k_3}\|$
with entries $k_i$ of Eq. (\ref{eq:km}),
we arrive at the second-order Lagrangian
\begin{equation}
\label{Leffgen}
L^{(2)}(\phi, \chi)=\frac{1}{2}\int d^3x\left[\sum_{i=1}^{3}\dot{\phi}_i^2+
\sum_{i,j =1}^{3}\dot{\chi}_iG^{-1}_{ij}\dot{\chi}_j
  - V(\phi, \chi)\right] -
   \theta\int d^3x\,\,Q^{(2)}(\phi, \chi)\,,
\end{equation}
with all $\theta$ dependence gathered in the reduced topological
charge density
\begin{equation} \label{eq:q2}
{Q}^{(2)}\, =\frac{1}{8\pi^2}\sum_{i=1}^3\left( \dot{\phi}_i
\beta_i +\sum^{a,b,c}_{cyclic}\dot{\chi}_i
M^{T}_{ia}(\phi_b-\phi_c)\,b_a\right)\,.
\end{equation}
The $Q^{(2)}$ in the effective Lagrangian (\ref{Leffgen})
can be represented as the divergence
\begin{equation}
\label{eq:q2s}
Q^{(2)} = \partial^\mu K^{(2)}_\mu
\end{equation}
of the four-vector $K^{(2)}_\mu=(K^{(2)}_0,K^{(2)}_i)$, with the components
\begin{eqnarray}
\label{K0un}
K^{(2)}_0 &=& \frac{1}{16\pi^2}
\sum_{cyclic}^{a,b,c} \left[(\phi_a-\phi_b)^2\Gamma_{acb}
-\frac{2}{3}\ g\ \phi_a\phi_b\phi_c\right],\\
K^{(2)}_i &=&\,\frac{1}{16\pi^2}\sum_{cyclic}^{a,b,c}
R^T_{ia}(\phi_b-\phi_c)^2\Gamma_{b0c},
\end{eqnarray}
with the space components of $\Gamma$ given in Eq.
(\ref{Gammaspace}), and the time components correspondingly defined  as
\begin{equation}
\Gamma_{a0b} = \left(\dot{ R} R^T \right)_{ab}\, .
\end{equation}
This completes our construction of the second-order
Lagrangian with all $\theta$ contributions gathered in a total
differential (\ref{eq:q2}) (see also Appendix D).
We have found the unconstrained analogue of the Chern-Simons current
$K_\mu^{(2)}$, linear in the derivatives.
Under the assumption that the vector part  $K^{(2)}_i$ vanishes
at spatial infinity, the unconstrained form of the Pontryagin index $p_1$
can be represented as the difference of the two surface integrals
\begin{equation}
\label{eq:achsgen}
W_\pm = \int d^3x\,\, K_0^{(2)}\left( t \to  \pm\infty, \, {\vec x}\right)\,,
\end{equation}
which are the winding number functional (\ref{clctr1}) for the physical
field $S$ in terms of principal-axes variables (\ref{eq:mainax}) at
$t \to\pm\infty $, respectively,
since $K_0^{(2)}(\phi,\chi)$ of Eq. (\ref{K0un}) coincides with the full
$K_0[S[\phi,\chi]]$ of Eq. (\ref{CSC}).
In the next section we shall show how for certain field configurations
it reduces to the Hopf number of the
mapping from the three-sphere $\mathbb{S}^3$ to the unit two-sphere
$\mathbb{S}^2$.

%%%%%%%%%%%%%%%%%%%%%%%%%%%%%%%%%%%%%% 6 %%%%%%%%%%%%%%%%%%%%%%%%%%%%%%%%%%%%%%

\section{Unconstrained theory for degenerate configurations  }

%%%%%%%%%%%%%%%%%%%%%%%%%%%%%%%%%%%%%%%%%%%%%%%%%%%%%%%%%%%%%%%%%%%%%%%%%%%%%%%%

The previous study  was restricted to
consideration of the domain of configuration space with $\det \|S\| \neq 0$,
where the change of variables (\ref{eq:gpottr}) is well defined.
In this section we would like to discuss the dynamics on
the special degenerate stratum (DS) with $\mbox{rank}\|S\| = 1$,
corresponding to the case of two eigenvalues of the matrix $S$ vanishing.
To investigate the dynamics on degenerate orbits it is in principle
necessary to use a decomposition of the gauge potential
different from our representation (\ref{eq:gpottr})
and the corresponding subsequent principal-axes transformation (\ref{eq:mainax}).
Instead of this, we shall use here the fact that the
degenerate orbits can be regarded as the boundary of the nondegenerate ones
and find the corresponding dynamics by taking the corresponding limit from
the nondegenerate orbits.
Assuming the validity of such an approach
we shall analyze the limit when
two eigenvalues of the symmetric matrix $S$  tend to zero.
\footnote{
%-----------------------------------------------------------
It can easily be checked that the degenerate stratum with
$\mbox{rank}\|S\|=1$ is dynamically invariant.
Furthermore, it is obvious from the representation
(\ref{eq:uncz}) of the unconstrained Hamiltonian that it is necessary to have
$\xi_k\rightarrow 0$ for some fixed $k$,
in order to obtain a finite contribution of the
kinetic term to the Hamiltonian in the limit
$\phi_i,\phi_j\rightarrow 0$ for $(i, j \neq k)$.}
%-----------------------------------------------------------------
Due to the cyclic symmetry under permutation of the diagonal fields
it is enough to choose one  singular configuration
\begin{equation}
\label{eq:conf}
\phi_1(x) = \phi_2(x) = 0\ \ \ {\rm and} \qquad \phi_3(x)
\ \quad {\rm arbitrary}\,.
\end{equation}
Note that for the configuration (\ref{eq:conf}) the spatially homogeneous part
(\ref{Vhom}) of the
square of the magnetic field vanishes and the potential term in the
Lagrangian (\ref{Leffgen}) reduces to the expression
\begin{eqnarray}
V =
&& \phi_3^2\big[(\Gamma_{2 1 3})^2+(\Gamma_{2 2 3})^2
           +(\Gamma_{2 3 3})^2
           +(\Gamma_{3 1 1})^2+(\Gamma_{3 2 1})^2
           +(\Gamma_{3 3 1})^2 + (\Gamma_{3 [12]})^2 \big]\nonumber\\
&& +\big[(X_1\phi_3)^2+(X_2\phi_3)^2\big]
   +2\phi_3\big[\Gamma_{3 3 1} X_1\phi_3
                   +\Gamma_{3 3 2} X_2\phi_3\big],
\label{V21}
\end{eqnarray}
which can be rewritten as \cite{KP,ErrKP}
\begin{equation}
\label{V22}
V= (\nabla \phi_3)^2
+ \phi_3^2\left[(\partial_i{\mathbf{n}})^2+ (\mathbf{n}\cdot
{\rm rot}{\ \mathbf{n}})^2\right]
 -(\mathbf{n} \cdot \nabla \phi_3)^2
+ ([\mathbf{n} \times \mbox{rot\ } \mathbf{n} ] \cdot \nabla \phi_3^2),
\end{equation}
introducing the unit vector
\begin{equation}
n_i(x):=R_{3i}[\chi(x)]\,.
\end{equation}
Hence the unconstrained second-order Lagrangian
corresponding to the degenerate stratum   with $\mbox{rank}\|S(x)\| = 1$
takes the form of the nonlinear $\sigma$-model type Lagrangian
\begin{eqnarray}
\label{eq:DSL}
L_{\rm DS} &=&
{1\over 2}\int d^3x
\Big[(\partial_\mu \phi_3)^2+  \phi_3^2(\partial_\mu \mathbf{n})^2-
 \phi_3^2(\mathbf{n}\cdot{\rm rot}{\ \mathbf{n}})^2
 +(\mathbf{n} \cdot \nabla \phi_3)^2\nn\\
&&\ \ \ \ \ \ \ \ \ \ \ \ \
- ([\mathbf{n} \times \mbox{rot\ } \mathbf{n} ] \cdot \nabla \phi_3^2)\Big]
-\theta\int d^3x\,\, Q_{\rm DS}
\end{eqnarray}
for the unit vector $\mathbf{n}(x)$ field  coupled to the field $\phi_3(x)$.
The density of the topological term $Q_{\rm DS}$ in the Lagrangian (\ref{eq:DSL})
can be represented as the divergence
\begin{equation}
Q_{\rm DS} = \partial_\mu K_{\rm DS}^\mu\,
\end{equation}
of the four-vector
\begin{equation}
K_{\rm DS}^\mu = \frac{1}{16\pi^2}\phi_3^2
\left([\mathbf{n}(x)\cdot \mbox{rot}\ \mathbf{n}(x)],
[\mathbf{n}(x)\times \dot{\mathbf{n}}(x)] \right).
\end{equation}
If we impose the usual boundary condition that the field ${\mathbf{n}}$
becomes time independent at spatial infinity,  the contribution from the
vector part  $K_{\rm DS}^i$ vanishes and
the unconstrained form of the Pontryagin topological index $p_1$
for the degenerate stratum with $\mbox{rank}\|S\|=1$ can be represented
as the difference
\begin{equation}
p_1 = n_+  - n_-\,
\end{equation}
of the surface integrals
\begin{equation}
\label{eq:achs}
n_{\pm} = \frac{1}{16\pi^2}\int d^3x
\left[ {\mathbf{V}}_{\pm}(\vec{x})\cdot \mbox{rot}\
{\mathbf{V}}_{\pm} (\vec{x}) \right]\,
\end{equation}
of the fields
\begin{equation}
{\mathbf{V}}_{\pm}(\vec{x}) := \lim_{t \to \pm \infty}\ \phi_3(x){\mathbf{n}}.
\end{equation}
We shall show now that the surface integrals (\ref{eq:achs}) are
Hopf invariants in the representation of Whitehead \cite{Whitehead}.

Under the Hopf mapping of a three-sphere to a two-sphere having unit radius,
$N\, : \mathbb{S}^3 \to \mathbb{S}^2$, the preimage of a point on
$\mathbb{S}^2$ is a closed loop.
The number $Q_H$ of times the loops corresponding to two distinct points
on $\mathbb{S}^2$ are linked to each other is the so-called Hopf invariant.
According to Whitehead \cite{Whitehead}, this linking number can be
represented by the integral
\begin{equation} \label{eq:Hopf}
Q_H = \frac{1}{32\pi^2}\int_{S^3}  w^1 \wedge w^2\,,
\end{equation}
with the so-called Hopf two-form curvature  $w^2 = H_{ij} dx^i  \wedge dx^j $
given in terms of the map $N$ as
\begin{equation}
\label{Hij1}
H_{ij}  = \varepsilon_{abc}N_a
\left( \partial_i N_b\right) \left(\partial_j N_c \right),
\end{equation}
and the one-form $w^1$ related to it via $w^2 =dw^1$.
Since the curvature $H_{ij}$ is divergence-free,
\begin{equation}
\varepsilon_{ijk}\partial_{i}H_{jk}=0,
\end{equation}
it can be represented as the rotation
\begin{equation}
\label{Hij2}
H_{ij} =\partial_i \mathcal{A}_j - \partial_j\mathcal{A}_i
\end{equation}
in terms of some vector field $\mathcal{A}_i$ ($i=1,2,3$)
defined over the whole of $\mathbb{S}^3$.
Thus the Hopf invariant takes the form
\begin{equation}
\label{eq:rot}
Q_{H} = \frac{1}{16\pi^2}\int d^3x
\left( {\mathbf{\mathcal{A}}}\cdot \mbox{rot}\
{\mathbf{\mathcal{A}}} \right).
\end{equation}

Therefore, the surface integrals (\ref{eq:achs}) are just Hopf invariants
in the Whitehead representation (\ref{eq:rot})
and the unconstrained form of the topological term $Q^{(2)}$
is a three-dimensional Abelian Chern-Simons term \cite{Jackiw}
with ``potential'' $V_i\ $ and the corresponding ``magnetic
field" $ \mbox{rot}\mathbf{V}$.
The topological term in the original $SU(2)$ Yang-Mills theory
reduces for rank-1 degenerate orbits not to a winding number,
but to the linking number $Q_H$ of the field lines.

We would like to end this section with two important open questions
to be posed for future investigations.
First, it would be very interesting to work out whether the classical
unconstrained theory obtained for degenerate field configurations can be used
to obtain some effective quantum model relevant
to the low-energy region of Yang-Mills theory, such as those proposed
and discussed recently in \cite{FaddeevNiemi}-\cite{Battye:1998}.
Second, due to the noncovariance of the symmetric gauge imposed,
the Lorentz transformation properties of the fields $\phi_3$ and
${\mathbf n}$ are nonstandard
(see, e.g., similar discussions for the case of the Coulomb gauge in
electrodynamics \cite{BjorkenDrell,HansonReggeTeitelboim,PavelPervushin}).
A careful investigation is necessary, taking into account surface contributions
to the unconstrained form of the generators of the Poincar\'{e} group.

%%%%%%%%%%%%%%%%%%%%%%%%%%%%%%%%%%%%%% 7 %%%%%%%%%%%%%%%%%%%%%%%%%%%%%%%%%%%%%%

\section{Conclusions and remarks}

%%%%%%%%%%%%%%%%%%%%%%%%%%%%%%%%%%%%%%%%%%%%%%%%%%%%%%%%%%%%%%%%%%%%%%%%%%%%%%%%

We have generalized the Hamiltonian reduction of
$SU(2)$ Yang-Mills gauge theory to the case of nonvanishing $\theta$ angle,
and shown that there is agreement between the reduced
and original constrained equations of motions.
We have employed an improved derivative expansion of the nonlocal
kinetic term in the unconstrained Hamiltonian obtained and
investigated it in the long-wavelength approximation.
The corresponding second-order Lagrangian has been constructed,
with all $\theta$ dependence gathered in the four-divergence of
a current, linear in the derivatives, which is the
unconstrained analogue of the original Chern-Simons current.

For the degenerate gauge field configurations $S$ with $\mbox{rank}\|S\| =1$,
we have argued that the long-wavelength Lagrangian obtained reduces to a
classical theory with an Abelian Chern-Simons term originating from the
Pontryagin topological functional.
Therefore the topological characteristic of the degenerate
configuration is given not by a winding number,
but by the linking number of the field lines.

Finally, let us comment on the Poincar\'{e} covariance of our
unconstrained version of Yang-Mills theory. It is well known that the
Hamiltonian formulation of degenerate theories reduced with the help of
noncovariant gauges destroy the manifest Poincar\'{e} invariance.
Our ``symmetric'' gauge condition (\ref{symgauge})
is not covariant under standard Lorentz transformations.
This, however, does not necessarily violate
the Poincare\'{e} invariance of our reduced theory.
Such a situation can be found in classical electrodynamics.
After imposing the Coulomb gauge condition the vector potential
ceases to be an ordinary Lorentz vector and
transforms nonhomogeneously under Lorentz transformations.
The standard Lorentz boosts are compensated by some additional
gauge-type transformation depending on the
boost parameters and the gauge potential itself
(see, e.g., \cite{BjorkenDrell,HansonReggeTeitelboim,PavelPervushin}).
As for the case of the Coulomb gauge in
electrodynamics, a thorough analysis of the Poincar\'{e} group representation
for our reduced theory obtained by imposing the symmetric gauge condition
is required. This problem is technically highly difficult and demands
special consideration that is beyond the scope of the present article.

%%%%%%%%%%%%%%%%%%%%%%%%%%%%%%%%%%%%%% 8 %%%%%%%%%%%%%%%%%%%%%%%%%%%%%%%%%

\section*{Acknowledgments}

%%%%%%%%%%%%%%%%%%%%%%%%%%%%%%%%%%%%%%%%%%%%%%%%%%%%%%%%%%%%%%%%%%%%%%%%%%%%

We are grateful for discussions with Z. Aouissat, R.~Horan,
A.~Kovner, A.~N.~Kvinikhidze, M.~Lavelle, M. D.~Mateev, D.~McMullan,
P.~Schuck, D. V.~Shirkov, M.~Staudacher and  A. N.~ Tavkhelidze.
A.K. and D.M. would like to thank Professor G. R\"opke
for his kind hospitality in the group ``Theoretische Vielteilchenphysik''
at the Fachbereich Physik of Rostock University, where part of this work was
done. A.K. thanks the Deutsche Forschungsgemeinschaft and
H.-P.P. the Bundesministerium fuer Forschung und Technologie for
financial support.

%%%%%%%%%%%%%%%%%%%%%%%%%%%%%%%%%%%%%% Appendix A %%%%%%%%%%%%%%%%%%%%%%%%%%%%%

\section*{Appendix A: Conventions and notation}

%%%%%%%%%%%%%%%%%%%%%%%%%%%%%%%%%%%%%%%%%%%%%%%%%%%%%%%%%%%%%%%%%%%%%%%%%%%%%%%%

\label{ap:A}

In this appendix, we collect the notation and definitions for $SU(2)$
Yang-Mills theory used in the text following \cite{Jackiw}.

The classical Yang-Mills action of the $su(2)$-valued connection one-form $A$
in four-dimensional Minkowski space-time with a metric $\eta =
\mbox{diag}\|1,-1,-1,-1\|$
reads
\begin{equation}
\label{eq:action}
I  =  - \frac{1}{g^2} \, \int  \, \mbox{tr} \, F \wedge {}^\ast\! F -
\, \frac{\theta}{8\pi^2\,g^2} \, \int \, \mbox{tr} \, F \wedge F \,,
\end{equation}
with the curvature two-form
\begin{equation}
F = d A + A \, \wedge\, A\,
\end{equation}
and its Hodge dual ${}^\ast\! F$.
The trace in Eq. (\ref{eq:action}) is calculated in the
anti-Hermitian $su(2)$ algebra basis
$\tau^a = \sigma^a/2\, i $ with Pauli matrices $\sigma^a, \, a=1,2,3$,
satisfying $[\tau_a,\, \tau_b] = \varepsilon_{abc}\,\tau_c$ and
$\mbox{tr}\left(\tau_a\tau_b\right)= - \frac{1}{2}\, \delta_{ab}$.

In the coordinate basis the components of the connection one-form $A$ are
\begin{equation}
A = g \, \tau^a \, A_\mu^a \, dx^\mu\,,
\end{equation}
and the components of the curvature two-form $F$ are
\begin{eqnarray}
F \,& = &\, \frac{1}{2}\, g\, \tau^a\, F^a_{\mu\nu}\, dx^\mu\,
\wedge\, dx^\nu\,,\\
F^a_{\mu\nu}& = &
\partial_\mu A^a_\nu - \partial_\nu A^a_\mu + g \,
\varepsilon^{abc} A^b_\mu A^c_\nu \,.
\end{eqnarray}
Its dual ${}^\ast\! F$ is given as
\begin{eqnarray}
{}^\ast\! F\, & = &\,
\frac{1}{2}\, g \, \tau^a\, {}^\ast\! F^a_{\mu\nu}\, dx^\mu\wedge dx^\nu\,,\\
{}^\ast\! F^a_{\mu\nu}\, &=&\, \frac{1}{2}\, \varepsilon_{\mu\nu\rho\sigma}
F^{a\, \rho\sigma}\,,
\end{eqnarray}
with the totally antisymmetric Levi-Civit\'{a} pseudotensor
$\varepsilon_{\mu\nu\rho\sigma}$, using the convention
\begin{equation}
\varepsilon^{0123} = -\, \varepsilon_{0123} = 1.
\end{equation}
The $\theta$ angle enters the classical action as the coefficient in front of
the Pontryagin index density
\begin{equation}
Q\, = -\, \frac{1}{8 \pi^2}\, \mbox{tr} \, F \wedge F.
\end{equation}
The Pontryagin index density is a closed form $dQ=0$ and thus
locally exact
\begin{equation}\label{eq:topcharge}
Q \, = \, dC,
\end{equation}
with the Chern three-form
\begin{equation}
C = - \frac{1}{8 \pi^2}\, {\rm tr}\left( A\wedge dA +
\frac{2}{3}\, A\wedge A\wedge A \right)\,.
\end{equation}
The corresponding Chern-Simons current $K^\mu$
is a dual of the three-form $C$,
\begin{equation}
K^\mu =
(1/3!)\, \varepsilon^{\mu\nu\rho\sigma}\, C_{\nu\rho\sigma}\,=
-\,\frac{1}{16 \pi^2}\, \varepsilon^{\mu\alpha\beta\gamma}\mbox{tr}
\left(
F_{\alpha\beta}\, A_\gamma - \frac{2}{3}\, A_\alpha A_\beta A_\gamma
\right).
\end{equation}
with the notations $A_\mu :=g \, \tau^a \, A_\mu^a \,$ and
$ F_{\mu\nu} :=g \, \tau^a \, F_{\mu\nu}^a \,$.
The chromomagnetic field is given by
\begin{equation}
B^a_{i} = \frac{1}{2} \, \varepsilon_{ijk}\, F^a_{jk}\, =
\varepsilon_{ijk}\, \left(\partial_j A_{ak} + \frac{g}{2}\,
\varepsilon_{abc}\, A_{bj}\, A_{ck}\right),
\end{equation}
and the covariant derivative in the adjoint representation as
\begin{equation}
\left( D_i (A) \right)_{ac}  =
\delta_{ac} \partial_i +  g\, \varepsilon_{abc} A_{bi}.
\end{equation}
Finally, we frequently use the matrix notation
\begin{equation}
A_{ai} := A^a_{i}\,,\qquad
B_{ai} := B^a_{i}\,.
\end{equation}

%%%%%%%%%%%%%%%%%%%%%%%%%%%%%%%%%%%%% Appendix B %%%%%%%%%%%%%%%%%%%%%%%%%%%%%%

\section*{Appendix B: On the existence of the ``symmetric gauge'' }

%%%%%%%%%%%%%%%%%%%%%%%%%%%%%%%%%%%%%%%%%%%%%%%%%%%%%%%%%%%%%%%%%%%%%%%%%%%%%%%%
\label{ap:B}

In this appendix we discuss the condition under which the symmetric gauge
\begin{equation}
\label{eq:symmgauge}
\chi_a (A) = \varepsilon_{abi}\,A_{bi}(x) = 0
\end{equation}
exists.

According to the conventional  gauge-fixing method
(see, e.g., \cite{FadSlav}), a gauge $\chi_a (A) = 0$ exists
if the corresponding equation
\begin{equation}
\label{eq:gf}
\chi_a (A^\omega) =0
\end{equation}
in terms of the gauge transformed potential
\begin{equation}
\label{eq:gtr}
A^\omega_{ai}\tau_a =
U^+ (\omega)
         \left( A_{ai}\tau_a
         + \frac{1}{g}\frac{\partial}{\partial x_i} \right)U(\omega)
\end{equation}
has a unique solution for the unknown function $\omega(x)$.\footnote{
%________________________________________________________________________________%
Here we assume that the second gauge condition $A_{a0}=0$ is satisfied and
the function $\omega(x)$ therefore depends only on the space coordinates.}
%________________________________________________________________________________%

Hence the symmetric gauge (\ref{eq:symmgauge}) exists if any gauge potential
$A$ can be made symmetric by a unique time-independent gauge transformation.
The equation that determines the gauge transformation $\omega(x)$ which converts
an arbitrary gauge potential $A(x)$ into its symmetric counterpart
can be written as a matrix equation
\begin{equation}
\label{eq:symgeq}
O^T(\omega)A-A^TO(\omega) = \frac{1}{g}\left[\Sigma(\omega)
                           -\Sigma^T(\omega)\right]\,,
\end{equation}
with the orthogonal $3\times 3$ matrix related to the $SU(2)$ group element
\begin{equation}\label{eq:ort}
O_{ab}(\omega) =
- 2\ \mbox{tr} \left[U^+(\omega)\tau_a U(\omega)\tau_b\right]
\end{equation}
and the $3\times 3$ matrix $\Sigma$
\begin{equation}
\label{eq:Omega}
\Sigma_{ai}(\omega):= - \frac{1}{4 i}
\ \varepsilon_{amn}\left(O^T(\omega)
\frac{\partial O(\omega)}{\partial x_i}\right)_{mn}\,.
\end{equation}

We shall now prove the following

{\it Theorem}.
For any nondegenerate matrix $A$
Eq. (\ref{eq:symgeq}) admits a unique solution
in the form of a $1/g$ expansion
\begin{equation}
\label{eq:1gsol}
O(\omega) = O^{(0)}\left[1\ +\sum_{n=1}^{\infty}
\left(\frac{1}{g}\right)^n X^{(n)}\right].
\end{equation}

{\it Proof}.
In order to prove the statement, we first note that equating coefficients
of equal powers in $1/g$ in the orthogonality condition $O^TO = OO^T=I$ of
the matrix $O$ imposes the condition of orthogonality of $O^{(0)}$,
\begin{equation}
\label{orth}
O^{(0)T}O^{(0)}=O^{(0)}O^{(0)T}=I,
\end{equation}
as well as the conditions
\begin{eqnarray}
&& X^{(1)} +  {X^{(1)}}^T = 0\,, \nn \\
&& X^{(2)} +  {X^{(2)}}^T +  X^{(1)}{X^{(1)}}^T = 0\,, \nn \\
&& \cdots  \quad\qquad  \cdots \,, \nn \\
&& X^{(n)} + {X^{(n)}}^T + \sum_{i+j=n}X^{(i)}{X^{(j)}}^T = 0\,, \nn \\
&& \cdots  \quad \qquad \cdots
\label{eq:ofn}
\end{eqnarray}
for the unknown functions $X^{(n)}$.
Furthermore, plugging expansion (\ref{eq:1gsol})
into Eq. (\ref{eq:symgeq}) and combining the terms of equal powers of $1/g$,
we find that the orthogonal matrix $O^{(0)}$ should satisfy Eq.
(\ref{eq:symgeq}) to leading order in $1/g$,
\begin{equation}
\label{eq:f0}
{O^{(0)}}^TA - A^T{O^{(0)}}=0\,,
\end{equation}
and the $X^{(n)}$ should satisfy the infinite set of equations
\begin{eqnarray}
&& {X^{(1)}}^T{O^{(0)}}^TA - A^T{O^{(0)}}{X^{(1)}}
= \Sigma^{(0)}- {\Sigma^{(0)}}\,, \nn\\
&& \cdots  \quad\qquad  \cdots \,, \nn \\
&& {X^{(n)}}^T{O^{(0)}}^TA- A^T{O^{(0)}}^T{X^{(n)}}
= \Sigma^{(n-1)}- {\Sigma^{(n-1)}}^T\,, \nn \\
&& \cdots  \quad \qquad \cdots~,
\label{eq:fn}
\end{eqnarray}
where the corresponding $1/g$ expansion for the matrix $\Sigma(\omega)$
\begin{equation}\label{eq:sigex}
\Sigma(\omega) = \sum_{n=0}^{\infty}
\left(\frac{1}{g}\right)^n \Sigma^{(n)}
\end{equation}
has been used.
Note that in the expansion (\ref{eq:sigex})
the $n$th  order term $\Sigma^{(n)}$ is given in terms of
$O^{(0)}$ and $X^{(a)}$ with $ a = 1,\cdots ,n-1$.

From the structure of Eqs. (\ref{orth})-(\ref{eq:fn}) one can see that the
solution to Eq. (\ref{eq:symgeq}) reduces to an algebraic problem.
Indeed, the solution to the first, homogeneous equation (\ref{eq:f0})
is given by the polar decomposition for the arbitrary matrix $A$,
\begin{equation}
\label{eq:f0s}
O^{(0)}= A S^{(0)-1 }\,, \quad\quad S^{(0)}=\sqrt{A A^T}~.
\end{equation}
This solution is unique only if $\det \|A\| \neq 0$.
It follows from the well-known property that the polar decomposition is valid
for an arbitrary matrix $A$, but the orthogonal matrix $O^{(0)}$
is unique only for nondegenerate matrices \cite{Gantmacher}.

To proceed further we use this solution and Eqs.
(\ref{eq:ofn}) for unknown $X$ to rewrite the remaining equations
(\ref{eq:fn}) as
\begin{eqnarray}
&& {X^{(1)}}S^{(0)} + S^{(0)} {X^{(1)}}   = C^{(0)}\,, \nn\\
&& \cdots  \quad\qquad  \cdots \,, \nn \\
&& {X^{(n)}}S^{(0)} + S^{(0)} {X^{(n)}} = C^{(n-1)}\,, \nn \\
&& \cdots  \quad \qquad \cdots~,
\label{eq:fn2}
\end{eqnarray}
where the $n$th  order coefficient  $C^{(n)}$ is given in terms of
$O^{(0)}$ and $X^{(1)}, X^{(2)},\dots,  X^{(n-1)} $.

Thus, starting from the zeroth-order term, the higher-order
terms $X^{(n)}$ are given recursively as solutions
of matrix equations of the type $X S^{(0)}+ S^{(0)} X = C$ with a known
symmetric positive definite matrix
$S^{(0)} =\sqrt{A A^T}$ and matrix $C$, expressed in terms of
the preceding  $X^{(a)}$, $ a=1, \dots ,n-1 $.
The theory of such algebraic equations is well
elaborated (see, e.g., \cite{Lancaster,Gantmacher}).
In particular, Theorem 8.5.1 in \cite{Lancaster}
states that for matrix equations for unknown matrix $X$ of the type
$XA + BX = C$, there is a unique solution
if and only if  the matrices
$A$ and -$B$ have no common eigenvalues.
Based on this theorem one can conclude that the unique  solution
to Eqs. (\ref{orth})-(\ref{eq:fn}) and hence to our original problem
(\ref{eq:symgeq}) exists always for any nondegenerate matrix $A$.

It is necessary to emphasize that in order to prove
the existence and uniqueness of the representation (\ref{eq:gpottr})
it should be shown additionally to the above Theorem that the
corresponding symmetric matrix field $S$,
\begin{equation}\label{eq:sex}
S(x) = \sum_{n=0}^{\infty}
\left(\frac{1}{g}\right)^n S^{(n)}(x),
\end{equation}
is sign definite. Above, the positive definiteness has been shown
only for the zeroth-order term $S^{(0)}=\sqrt{A A^T}$.
The study of this problem,
as well as an analogous investigation for the degenerate field configurations $A$
with $\det \|A \|= 0$, are
beyond the scope of this appendix and will be discussed in detail elsewhere.
Here we limit ourselves to the consideration of a specific example,
elucidating the generic picture.

In the case that the matrix $A$ is degenerate, we encounter the problem
of Gribov's copies.
As an illustration of the nonuniqueness of the gauge transformation that
turns a given field configuration $A$ into the corresponding symmetric form,
we consider the ``degenerate'' field
\begin{equation}
\label{eq:wu-yang}
A_{a0} = 0\,, \qquad
A_{ai} = - \frac{1}{g r}\  \varepsilon_{aic}\hat{r}_c~,
\end{equation}
known as the non-Abelian Wu-Yang monopole field, with the unit vector
$\hat{r}_a = x_a/r\,$ and $\, r = \sqrt{x_1^2+x_2^2 + x_3^2}$\,.

Performing the gauge transformation
\begin{equation}
\label{eg:gtrw}
S_{ai}\tau_a= U^+ (\omega)
\left( A_{ai}\tau_a + \frac{1}{g}\frac{\partial}{\partial x_i} \right)U(\omega),
\end{equation}
with $U(\omega)= e^{\omega_a\tau_a}$ parametrized by one time-independent
spherical symmetric function
\begin{equation}\label{eq:fgtr}
\omega_a = f(r)\ \hat{r}_a\,,
\end{equation}
the Wu-Yang monopole configuration (\ref{eq:wu-yang}),
antisymmetric in space and color indices, can be brought into the
``symmetric form''
\begin{equation}
\label{eq:sinsy}
S_{ai}^{\pm} = \pm \frac{\sqrt{3}}{g r}
\left(\delta_{ai} - \hat{r}_a \hat{r}_i\right),
\end{equation}
if the function $f(r)$ is constant and takes four values:
\begin{equation}
\label{eq:wutr}
f(r) = \Bigg\{ \begin{array}{cc}\pi/3~,~ 7\pi/3\, &\quad {\rm for}\quad (+),
       \\ 5\pi/3~,~ 11\pi/3\, &\quad {\rm for}\quad (-). \end{array}
\end{equation}
Here $S^+$ can be obtained from the
Wu-Yang monopole configuration (\ref{eq:wu-yang}) by applying two different
gauge transformations with $\, f(r)=\pi /3\,, 7\pi/3$,
\begin{equation}\label{eq:gtrv_}
U_{1,2} =\pm \Big(\frac{\sqrt{3}}{2} - \hat{r}\cdot \tau \Big)\,,
\end{equation}
while the $S^{-}$ configuration
can be reached using $\, f(r)= 5\pi/3\,, 11\pi/3$,
\begin{equation}\label{eq:gtrv-}
U_{3, 4} =\mp \Big(\frac{\sqrt{3}}{2}+ \hat{r}\cdot \tau \Big)\,.
\end{equation}

Here it is in order to make the following comments.

For the above gauge transformations we have $\lim_{r\to \infty}U \neq \pm I$.
Thus they are neither small gauge transformations nor
large gauge transformations belonging to any integer $n$-homotopy class
\cite{Jackiw}.

The symmetric configurations (\ref{eq:sinsy}) corresponding to the
Wu-Yang monopole lie on the stratum of degenerate symmetric matrices
with one eigenvalue vanishing and two eigenvalues equal to each other.

The symmetric configurations $S^+$ and $S^-$ in Eq. (\ref{eq:sinsy})
with twofold Gribov degeneracy
are related to each other by parity conjugation.

%%%%%%%%%%%%%%%%%%%%%%%%%%%%%%%%%%%%%% Appendix C %%%%%%%%%%%%%%%%%%%%%%%%%%%%%%

\section*{Appendix C: Proof of
$\theta$ dependence of the naive $1/g$ approximation }

%%%%%%%%%%%%%%%%%%%%%%%%%%%%%%%%%%%%%%%%%%%%%%%%%%%%%%%%%%%%%%%%%%%%%%%%%%%%%%%%
\label{ap:C}

In this appendix it is shown that straightforward application of expansion
of the nonlocal part $P_a$ of the kinetic term in the unconstrained Hamiltonian
to zeroth order discussed in Sec. \ref{Sec:IV1}, leads to the
appearance of $\theta$ dependence of the reduced system on the classical level.
Expressing the Hamiltonian (\ref{eq:ham2}), in terms of the principal-axes
variables, defined in Sec. \ref{sec:V}, and performing an
inverse Legendre transformation, one obtains the Lagrangian density
\begin{eqnarray}
\label{Leffgent2}
{\cal L}^{(2)}(\phi, \chi)&=&\frac{1}{2}\left(\sum_{i=1}^{3}\dot{\phi}_i^2+
\sum_{i,j =1}^{3}\dot{\chi}_iG^{-1}_{ij}\dot{\chi}_j
  - V(\phi, \chi)\right)- \frac{1}{2}\left(\frac{\theta}{8\pi^2}\right)^2
  \sum_{cyclic}^{i,j,k}\frac{\Delta^2_i}{\phi^2_j+\phi^2_k}\nn\\ &&-
\frac{\theta}{8\pi^2}\sum_{a=1}^3\left[ \dot{\phi}_a \beta_a
+\sum_{cyclic}^{i,j,k}\dot{\chi}_a
M^{T}_{ai}(\phi_j-\phi_k)\left(b_i+
\frac{(\phi_j-\phi_k)}{\phi^2_j+\phi^2_k}\Delta_i\right)\right]\,,
\end{eqnarray}
denoting the difference
\begin{equation}
\Delta_i=\frac{1}{2}(\phi_j-\phi_k)b_i-(\phi_j+\phi_k)\,\sum_{s=1}^3
R_{is}B^{(-)}_s \quad (\mbox{cyclic permutations}\,\,i\neq j\neq k),
\end{equation}
with $b_i$ of Eq. (\ref{b_i}) and $B^{(-)}_i$ of Eq. (\ref{B-}),
or, explicitly,
\begin{eqnarray}
\Delta_i &=&
-\big[X_i(\phi_j\phi_k)+(\Gamma_{ijj}+\Gamma_{ikk})\phi_j\phi_k
-\phi_i(\phi_j\Gamma_{ikk}+\phi_k\Gamma_{ijj})\big]\nonumber\\
&& (\mbox{cyclic permutations}\,\,i\neq j\neq k).
\end{eqnarray}
It easy to convince ourselves that the term proportional to $\theta^2$
is not a surface term. Indeed, considering for simplicity configurations of
spatially constant angular variables $\chi_i$ and
$\phi_1=\phi_2=\phi_3=:\phi\,$, it reduces to
\begin{equation}
-\left(\frac{\theta}{8\pi^2}\right)^2\sum_{i=1}^{3}\,\,
\partial_i\phi\,\, \partial_i\phi~,
\end{equation}
which is not a four-divergence.
For $\Delta_i=0$ the Lagrangian density
(\ref{Leffgent2}) reduces to Eq. (\ref{Leffgen}), obtained
from the improved  Hamiltonian (\ref{eq:iham2}),
free of the divergence problem.

%%%%%%%%%%%%%%%%%%%%%%%%%%%%%%%%%%%%%% Appendix  %%%%%%%%%%%%%%%%%%%%%%%%%%%%%%

\section*{Appendix D: Representation of the
unconstrained fields in the basis of principal-axes variables}

%%%%%%%%%%%%%%%%%%%%%%%%%%%%%%%%%%%%%%%%%%%%%%%%%%%%%%%%%%%%%%%%%%%%%%%%%%%%%%%%
\label{ap:D}

Starting from the coordinate basis expression of $S$ in Eq. (\ref{3S}),
we observe that the principal-axes transformation (\ref{eq:mainax})
corresponds to the representation
\begin{equation}
%\label{eq:sof}
S =\sum_{a=1}^3 e_a\,\phi_{a} \,\omega_a \,,
\end{equation}
with the one-forms
\begin{equation}
\omega_i :=\sum_{j=1}^3 R_{ij}[\chi(x)]dx_j \,, \qquad i = 1,2,3
\end{equation}
and the $su(2)$ Lie algebra basis
\begin{equation}
e_a := \sum_{b=1}^3 R_{ab}[\chi(x)]\tau_b\,, \qquad a = 1,2,3~.
\end{equation}

\subsection*{D1. Unconstrained magnetic field}

The physical chromomagnetic fields $B_{ai}(S)$, given in Eq. (\ref{redB}),
can be regarded as the components of the dual ${}^\ast\! F^{(3)}$
\begin{eqnarray}
B_{ai}(S) =\frac{1}{2}\,\sum_{i,j=1}^3 \varepsilon_{ijk}\, F^{(3)}_{\ ajk}
\nonumber
\end{eqnarray}
of the curvature two-form $F^{(3)}$, defined
in terms of the symmetric one-form $S$ in Eq. (\ref{3F}) as
\begin{eqnarray}
F^{(3)} = d S + S\, \wedge\, S \nonumber
\end{eqnarray}
In the principal-axes basis the components of the non-Abelian field
strength $F^{(3)}$ read
\begin{equation}
%\label{eq:chrm}
F^{(3)}_{aij} =
\delta_{aj}\, X_i\, \phi_j - \delta_{ai}\, X_j\, \phi_i +
\phi_i\, \Gamma_{aji} - \phi_j \, \Gamma_{aij} +
\Gamma_{a[ij]}\, \phi_a +
g\varepsilon_{aij}\, \phi_i \,\phi_j \,,
\quad \mbox{(no summation)}\,,
\end{equation}
with the components of the connection one-form $\Gamma$ defined as
\begin{eqnarray}
%\label{Gammaspace}
\Gamma_{aib} := \left(X_i R\, R^T \right)_{ab}\,,\nonumber
\end{eqnarray}
and the vector fields
\begin{eqnarray}
X_i := \sum_{j=1}^3 R_{ij}\,\partial_j \nonumber
\end{eqnarray}
dual to the one-forms $\omega_j\,, \, \omega_i( X_j) = \delta_{ij}$,
and acting on the basis elements $e_a$ as
\begin{eqnarray}
X_i \, e_a = -\sum_{b=1}^3 \Gamma_{bia}e_b\,.
\end{eqnarray}
The explicit expressions for the diagonal components $\beta_i$ and
the off-diagonal components $b_i$ of the
symmetric part of the chromomagnetic field
\begin{equation}
\label{B+}
B^{(+)} = R^T(\chi)\,
 \sum_{i=1}^3\,\left( \beta_i\overline{\alpha}_i  +
\frac{1}{2}b_i\alpha_i \right)\, R(\chi)
\end{equation}
are given in terms of the diagonal fields $\phi_i$ and
the angular fields $\chi_i$ in cyclic form
\begin{eqnarray}
&&
\beta_i =
g\phi_j\phi_k - (\phi_i-\phi_j)\Gamma_{ikj} +
(\phi_i-\phi_k)\Gamma_{ijk}
\qquad (\mbox{cyclic permutations}\,\, i \neq j \neq k),\nonumber\\
&&
b_i=
X_i(\phi_j-\phi_k)-(\phi_i-\phi_j)\Gamma_{ijj} +
(\phi_i-\phi_k)\Gamma_{ikk}
\qquad (\mbox{cyclic permutations}\,\, i \neq j \neq k),\nonumber
%\label{b_i}
\end{eqnarray}
and the antisymmetric part $B^{(-)}_i$ of the unconstrained magnetic field is
\begin{equation}
\label{B-}
B^{(-)}_i=\frac{1}{2}\sum_{cyclic}^{a,b,c} R^T_{ia}\left[X_a(\phi_b+\phi_c)+
(\phi_b-\phi_a)\Gamma_{abb}+(\phi_c-\phi_a)\Gamma_{acc}\right] \,.
\end{equation}

\subsection*{D2. Unconstrained Chern-Simons three-form}

Using the Maurer-Cartan structure equations for the one-forms $\omega_i$
\begin{equation}
%\label{eg:streq}
d\omega_a  = \sum_{c=1}^3\Gamma_{a0c} dt\wedge \omega_c +
\sum_{b,c=1}^3\Gamma_{abc}\omega_b\wedge \omega_c~,
\end{equation}
with the space components of $\Gamma$ given in Eq.
(\ref{Gammaspace}), and the time components correspondingly defined  as
\begin{eqnarray}
\Gamma_{a0b} = \left(\dot{ R} R^T \right)_{ab}\,,\nonumber
\end{eqnarray}
Eq. (\ref{eq:q2})
can be written as
\begin{equation}
\label{dC}
Q^{(2)}= d C^{(2)}
\end{equation}
with the three-form
\begin{eqnarray}
 C^{(2)}&=& \frac{1}{8\pi^2}\sum_{a<b}^3 (\phi_a-\phi_b)^2\,\Gamma_{a0b}
dt\wedge\omega_a \wedge  \omega_b  \nn\\
&&\quad\quad\quad
-\frac{3}{8\pi^2}\sum_{cyclic}^{a,b,c} \left[(\phi_a-\phi_b)^2\,\Gamma_{acb}
-\frac{2}{3}\ \varepsilon_{abc}\phi_1\phi_2\phi_3\right]
\omega_a \wedge \omega_b \wedge \omega_c~.
\end{eqnarray}

%&&&&&&&&&&&&&&&&&&&&&&&&&&&&&&&&&&&&&&&&&&&&&&&&&&&&&&&&&&&&&&&&&&&&&&&&&&&%


\begin{thebibliography}{99}
\bibitem{ErrKP}
Erratum to \cite{KP}: The last line of Eq. (3.42) should be replaced by
$\left(\Gamma^1_{\ 23}\, \phi_3 + \Gamma^1_{\ 32}\, \phi_2 +
\Gamma^1_{\ [23]}\, \phi_1 - g\, \phi_2\, \phi_3\right)^2$.
Correspondingly, in formula (4.6) the term
$\left(\Gamma^3_{\ [12]}\, \phi_3\right)^2$
should be added.
Finally, in formulas (4.8), (4.10), and (4.11) the term
$\left({\mathbf n}\cdot \mbox{rot}\ {\mathbf n} \right)^2$
should be included.
The correct formulas are given in the present work in
Eqs. (\ref{Vinhom1}),(\ref{V21}), and (\ref{V22}) respectively.
\end{thebibliography}
\end{document}